\documentclass[a4paper,
              ]{jacow}
\makeatletter%
	\ifboolexpr{bool{xetex}}
	 {\renewcommand{\Gin@extensions}{.pdf,%
	                    .png,.jpg,.bmp,.pict,.tif,.psd,.mac,.sga,.tga,.gif,%
	                    .eps,.ps,%
	                    }}{}
\makeatother

\ifboolexpr{bool{xetex} or bool{luatex}} 
 {}                                      
 {\usepackage[utf8]{inputenc}}           

\usepackage[USenglish]{babel}			 

\usepackage[final]{pdfpages}
\usepackage{multirow}
\usepackage{ragged2e}

\ifboolexpr{bool{jacowbiblatex}}%
 {%
  \addbibresource{jacow-test.bib}
  \addbibresource{biblatex-examples.bib}
 }{}
\begin{document}

\title{Simulations of the Acceleration of Externally Injected Electrons in a Plasma Excited in the Linear Regime}

\author{Nicolas Delerue\thanks{delerue@lal.in2p3.fr}, Christelle Bruni, St\'ephane Jenzer, \\ LAL, Univ. Paris-Sud, CNRS/IN2P3, Universit\'e Paris-Saclay, Orsay, France.\\
Sophie Kazamias, Bruno Lucas, Gilles Maynard, Laboratoire de Physique des Gaz et des Plasmas, \\Univ. Paris-Sud, CNRS/INP, Universit\'e Paris-Saclay, Orsay, France.\\
Moana Pittman, CLUPS, Univ. Paris-Sud, Universit\'e Paris-Saclay, Orsay, France.\\}
\maketitle

\begin{abstract}
   We have investigated numerically the coupling between a 10 \si{MeV} electron bunch of high charge (\SI{> 100}{pc}) with a laser generated accelerating plasma wave. Our results show that a high efficiency coupling can be achieved using a \SI{50}{TW}, \SI{100}{\micro \meter} wide laser beam, yielding accelerating field above \SI{1}{ GV/m}. We propose an experiment where these predictions could be tested.
\end{abstract}

\section{Introduction}
Many important results have been reported over the past years about Laser-driven plasma accelerators~\cite{PhysRevLett.113.245002}, however one of the questions that remain unanswered 
is that of the acceleration of an electron beam in the few tens MeV range externally injected. At such energies,  the electron beam will be spread over several dizains of femtoseconds after few meters of drift space.
 Such duration is comparable to the typical period of plasma oscillations for a plasma with density of the order of \SI{e17}{cm^{-3}}.  This can be a limitation for external injection experiments as the electrons arriving at the wrong phase will not be captured and accelerated. One  solution to mitigate that effect is to  compress the electrons in time before accelerating them. This compression followed by acceleration is one of the main iters that we plan to investigate in  our experiment.

\subsection{Plasma acceleration in the linear regime}

A high power laser pulse sent in a low density hydrogen or helium gas can ionise it and create a wake characterized by strong electric and magnetic fields. If the pulse is not too intense ($I<\SI{e18}{W/cm^2}$), then the plasma will be weakly driven by the laser pulse, this is called the linear regime~\cite{Gorbunov:1987,Sprangle:1987}. 

In such conditions we can use or define the following quantities:
\begin{itemize}
 \item{Maximum Accelerating field} $E_0 = \frac{2 \pi m_e c^2}{e \lambda_p} $ hence  \\ $$E_0 [GV / m]= 96.2 \sqrt{n_e [\SI{e18}{cm^{-3}}]}$$.
 \item{Longitudinal accelerating field} \\ $$E_{0z} = \frac{\eta}{4}  a_0^2 \cos(k_p d_l) \exp(- \frac{2 \rho^2}{w_z^2}) \times E_0$$.
\item{Radial accelerating field}\\ $E_{0r} = \frac{\rho}{k_p w_z^2} \eta  a_0^2 \sin(k_p d_l) \exp(- \frac{2 \rho^2}{w_z^2}) \times E_0$
\end{itemize}
with $m_e$ the electron mass, $e$ the electron charge, $\lambda$ the laser wavelength (\SI{0.8}{\micro \meter}), $n_e$ the plasma density,  $\lambda_p$ the plasma wavelength ($\lambda_p = \lambda \times \sqrt{\frac{n_c}{n_e}}$) , $k_p$ the plasma wave number,  $\eta$ the laser-plasma coupling $a_0$ the plasma relativistic limit, $d_l$ the laser distance behind the pulse, $\rho$ the radial distance and $w_z$ the laser waist radius at position $z$ (and $w_0$ at $z=0$, the focal point).

Therefore a pressure of \SI{4e17}{cm^{-3}} will give a maximum longitudinal accelerating field of more than \SI{10}{GV/m}. This corresponds to a plasma wavelength of about \SI{50}{\micro m} (that is about \SI{180}{fs}). It is important to note also that the radial accelerating field can take either positive or a negative value, that is, it can be either focussing or defocussing.

With a \SI{2}{Joules} laser focused on a \SI{55}{\micro \meter} waist we get a Rayleigh length of about 1cm. This will give a sufficient length to compress and accelerate the electrons. These electrons must also be focussed in a comparable volume.

\subsection{Proposed experiment}

\begin{figure}[bthp]
    \centering
    \includegraphics*[width=\linewidth]{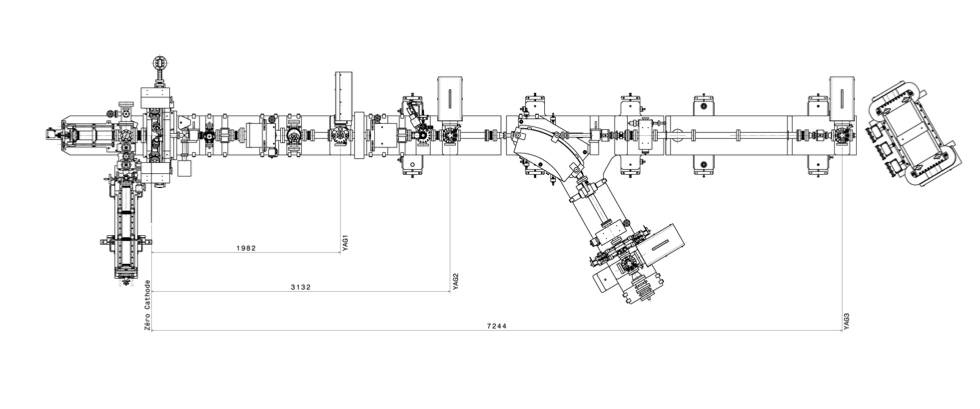}
     \caption{The PHIL beam line. The electrons are produced on the left of the image and travel toward the right. A spectrometer magnet can be seen in the middle of the beam line.}
    \label{PHIL}
\end{figure}

The proposed experiment will be performed within the ESCULAP  ({\em ElectronS CoUrts et LAsers Plasmas}) installation at LAL, by combining the LASERIX laser~\cite{Ple:07,Zimmer:10,Delmas:14}  with the PHIL photoinjector~\cite{1748-0221-8-01-T01001,Vinatier2015222}. PHIL is a conventional  \SI{5}{MeV} Photoinjector that is currently being upgraded to \SI{10}{MeV}. A layout of PHIL is shown on figure~\ref{PHIL}. Laserix is a 50 TW, 50~fs high-power Ti:Sa laser. As part of ESCULAP a small leak from Laserix will be directed on the PHIL photocathode to produce short electron pulses. Laser-driven plasma acceleration experiment will be performed by injecting simulaneously the high energy laser beam with the PHIL generated relativistic electron beam into a few cm gas cell.

\begin{figure}[htbp]
 \centering
  \includegraphics*[width=\linewidth]{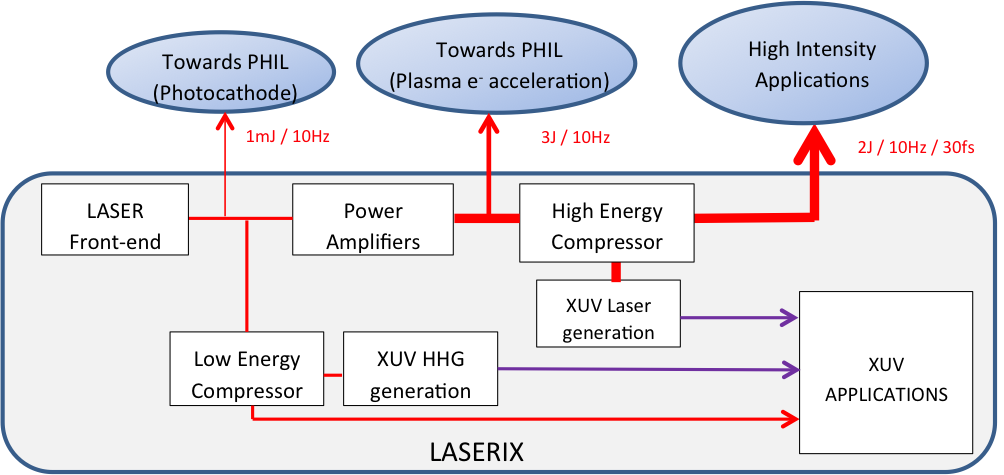}\\
  \caption{Scheme of the current LASERIX installation at LAL showing the new high intensity laser beamlines for emerging applications.}
   \label{fig:scheme_laserix_lal}
\end{figure}

\begin{figure}[htbp]
  \centering
  \includegraphics[width=0.9\linewidth]{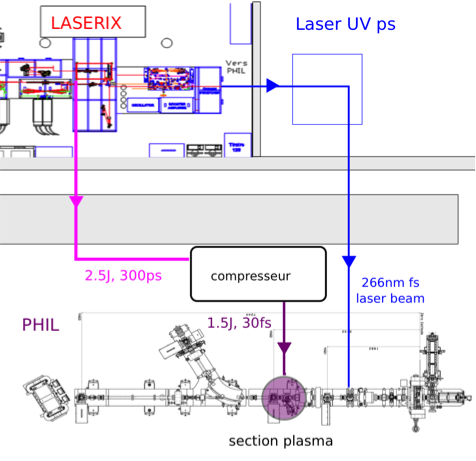}
  \caption{The PHIL and Laserix facilities. A small leak from Laserix will be sent on the PHIL photocathode and the reminder will be sent in the plasma chamber.}
  \label{PHIL_and_Laserix}
\end{figure}

\section{Simulations}

\begin{figure}[htbp]
  \centering
  \includegraphics[width=0.9\linewidth]{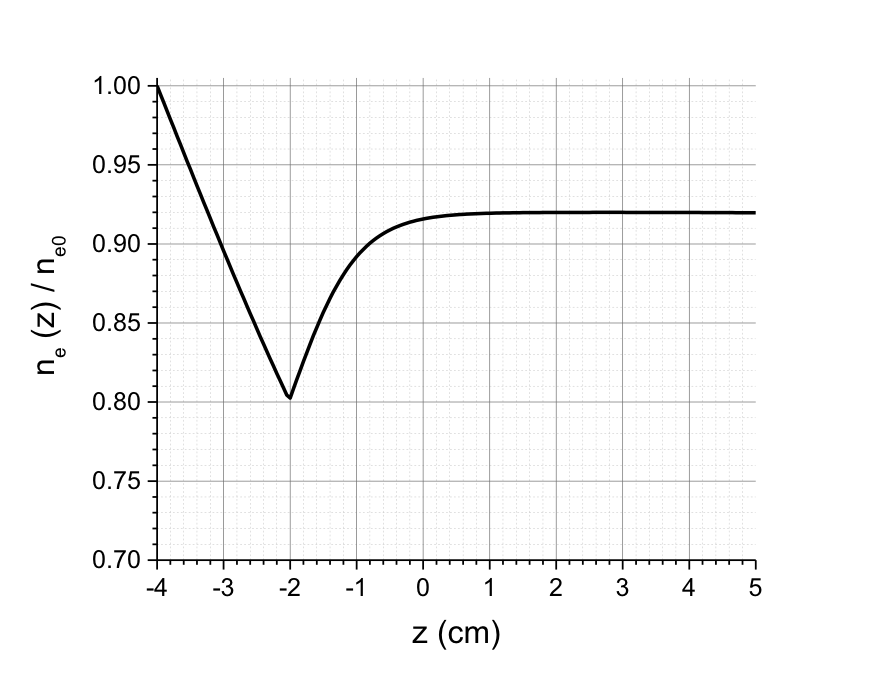}
  \caption{Density profile along the plasma axis. The maximum density, $n_{e0}$ is \SI{4e17}{cm^{-3}}.}
  \label{fig:plasma_density}
\end{figure}

For our simulations we have considered electrons with an energy of \SI{9.5}{MeV} focussed on a low density $H_2$ plasma. At the entrance of the plasma the electrons have a transverse size of \SI{170}{\micro \meter}, they are covering with an half-angle of \SI{3}{mrad} and the FWHM duration of the bunch\footnote{This is optimistic with respect to what has been achieved on PHIL so far but a short bunch production scheme is under study.} is \SI{75}{fs}. The total length of the plasma cell considered is \SI{9}{cm} but the cell has been designed so that it has a varying pressure according to the profile shown on figure~\ref{fig:plasma_density}.

The simultations were done using an adapted version of the numerical code WAKE-EP~\cite{WAKE}.

The aim of this special density profile is to achieve a radial and longitudinal compression of the electron bunch before its acceleration. The first part of the density profile (decreasing pressure gradient) will keep all the electron together in the focussing  phase of the plasma wake. As the electrons have a relatively low $\gamma$ the difference in accelerating gradient experienced between the head and the tail of the bunch will compress them all together. Once this is achieved the second part of the density profile (increasing pressure gradient) will keep the bunch together at the back of the wave to accelerate them with the highest field.

\begin{figure*}[tbp]
  \centering
  \vspace{-1cm}
  \begin{tabular}{cccc}
  (a) &  \includegraphics[width=0.4\linewidth]{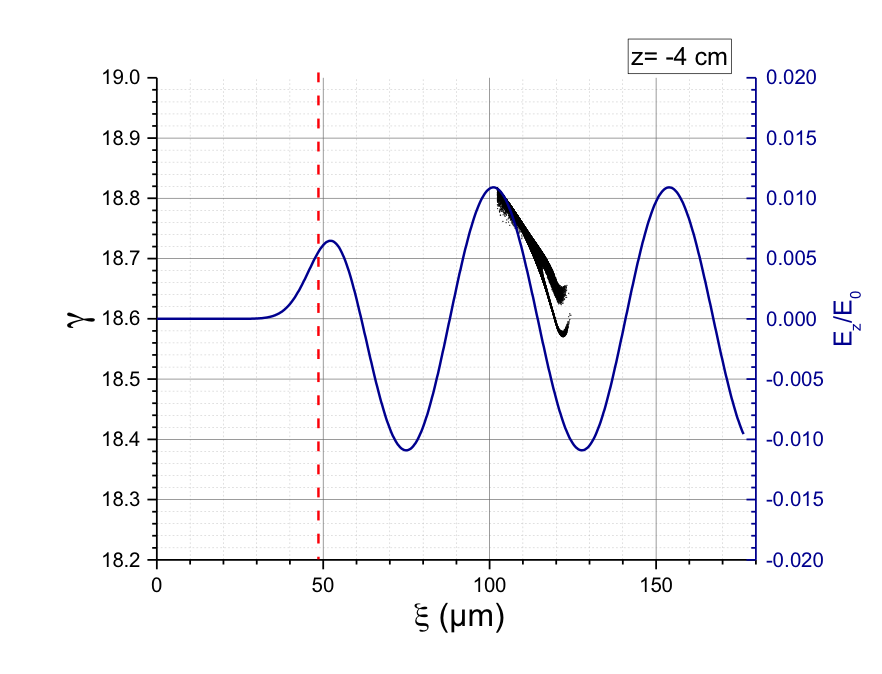} &
 (b) &  \includegraphics[width=0.4\linewidth]{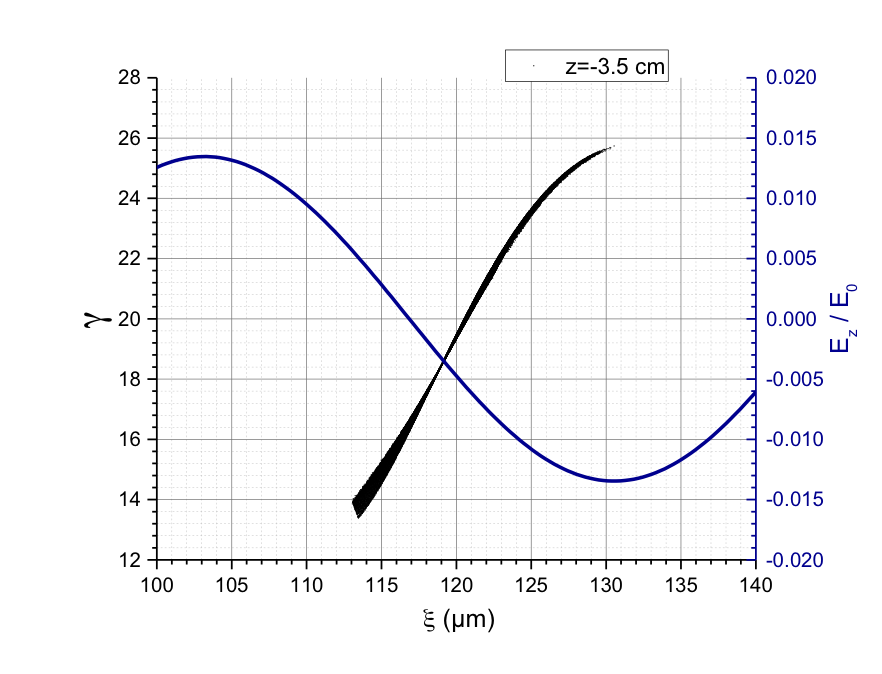} \\
 (c) & \includegraphics[width=0.4\linewidth]{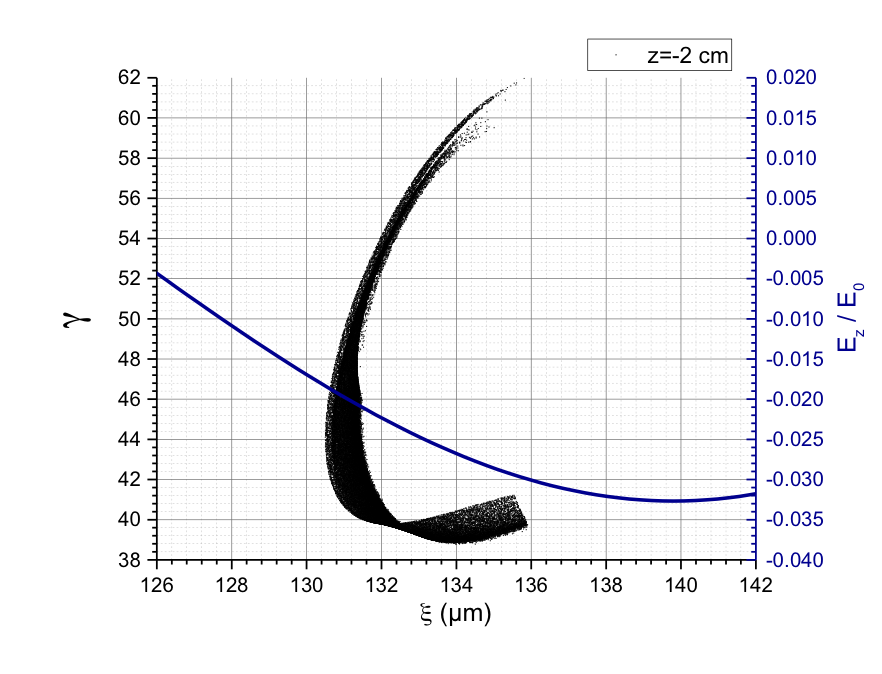} &
 (d) &  \includegraphics[width=0.4\linewidth]{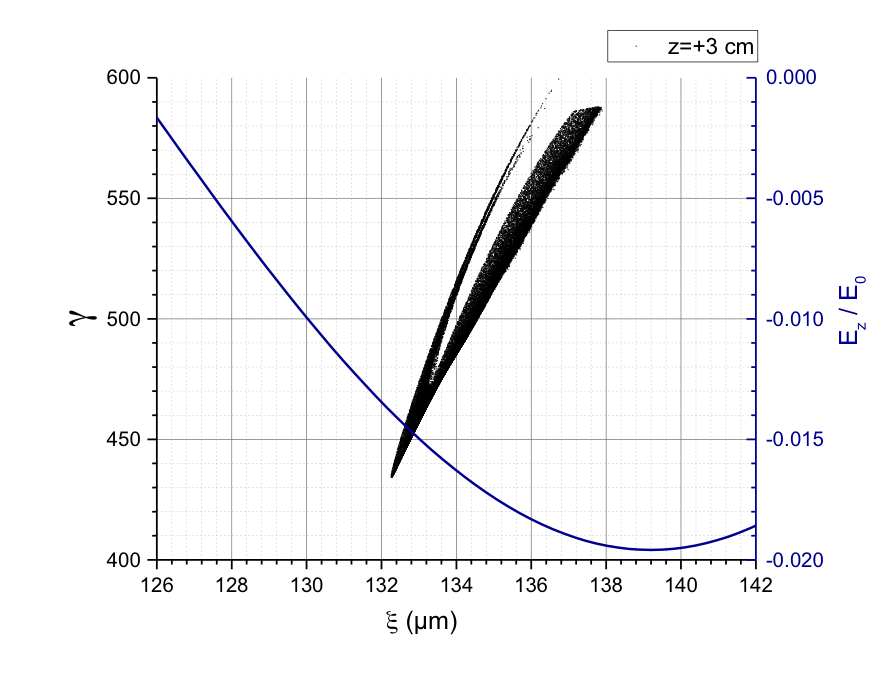} \\
  \end{tabular}
  \caption{Lorentz factor of the electrons (black and left vertical axis) and longitudinal accelerating field divided by $E_0 = mc \omega_p / e = \SI{608}{MV/cm}$ ( blue and right vertical axis), versus the distance behind the laser pulse at different z positions in the plasma. }
  \label{fig:simulations}
\end{figure*}

\begin{figure}[htbp]
  \centering
  \vspace*{-1cm}
  \includegraphics[width=0.95\linewidth]{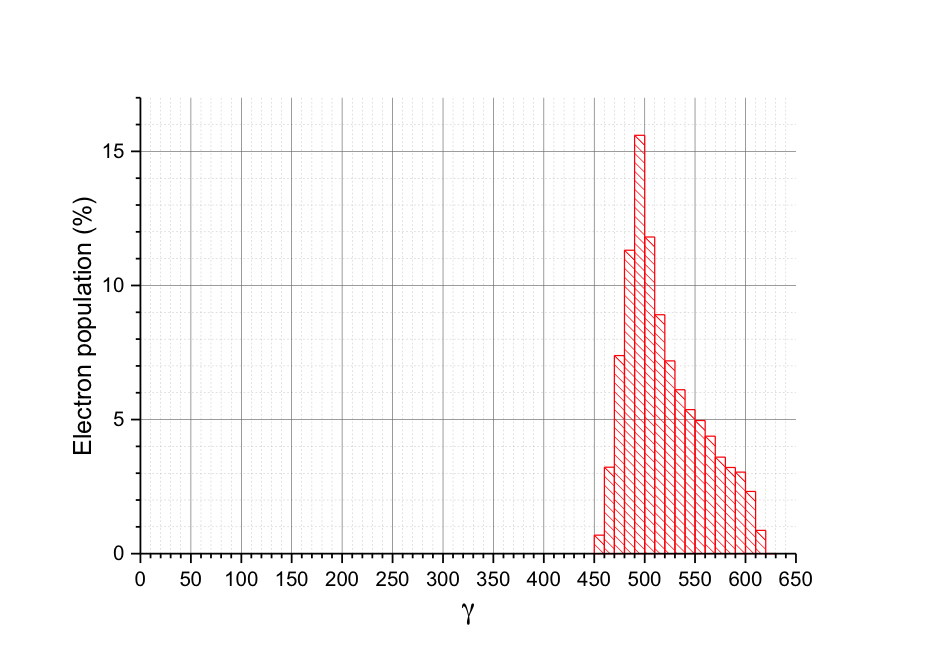}
  \caption{Energy distribution at the end of the acceleration process.}
  \vspace*{-0.6 cm}
  \label{fig:energy_dist}
  \end{figure}

On figure~\ref{fig:simulations} (a) one can see the distribution of the electrons (in black) and of the laser wake (in blue) at $z=\SI{-4}{cm}$, the entrance of the plasma cell. We can see that at injection the electron bunch (coming from a conventional accelerator simulated using ASTRA~\cite{astra}) have a large time spread and a small energy spread. As they progress through the decreasing gradient ramp the trailing electrons will experience a higher accelerating field than the electrons  at the front. As at these energy they are barely relativistic this difference will result in these electrons almost catching up with the leading one and the beam will get compressed in time. This is illustrated by figure~\ref{fig:simulations} (b-c). On figure~\ref{fig:simulations} (b) one can see that the distance between the leading and trailing electrons has significantly reduced and the trailing electrons have now more energy than the leading ones. On figure~\ref{fig:simulations} (c) the trailing electrons are even overtaking the leading ones and the bunch is compressed in only a few micrometers. It is important to note that this compression completely erases the initial energy spread of the bunch and its time spread. Once this process is over, after  $z=\SI{-2}{cm}$,  the increase in plasma density and in laser intensity will significantly accelerate the electrons. On figure~\ref{fig:simulations} (d) one can see that at the end of the accelerating process the electrons reach a Lorentz factor $\gamma$ of about 500 with slightly more than 15\% energy spread (figure~\ref{fig:energy_dist}).

\section{Conclusions}

External injection of electrons with an energy of about \SI{10}{MeV} in a low density plasma excited by a high power laser has the potential of bringing the energy of these electrons up \SI{250}{MeV}. This will require the use of a specially designed plasma cell capable of providing a two gradient ramp over a few centimeters. We plan to investigate this scheme experimentally using the ESCULAP facility installed on the Univ. Paris-Sud campus.





\end{document}